\documentclass[twocolumn,twoside,slac_two]{revtex4}

\usepackage{graphicx}
\usepackage{fancyhdr}
\begin{document}

\title{Three years of Fermi GBM Earth Occultation Monitoring: Observations of Hard X-ray/Soft Gamma-Ray Sources}

%

\author{P. Jenke}
\affiliation{University of Alabama in Huntsville, 301 Sparkman Drive, Huntsville, Alabama, USA}
\author{Colleen A. Wilson-Hodge}
\affiliation{ZP 12 Astrophysics Office, NASA Marshall Space Flight Center, Huntsville, AL 35812, USA}
\author{Gary L. Case}
\affiliation{Department of Physics, La Sierra University, 4500 Riverwalk Pkwy, Riverside, Ca 92505}
\author{Michael L. Cherry}
\affiliation{Department of Physics and Astronomy, Louisiana State University, Baton Rouge, LA,
70803, USA}
\author{James Rodi}
\affiliation{Department of Physics and Astronomy, Louisiana State University, Baton Rouge, LA,
70803, USA
}
\author{Ascension Camero-Arranz}
\affiliation{Instituto de Ciencias del Espacio (IEEC-CSIC), Campus UAB, Torre C5, 2a planta,
08193 Barcelona, Spain}
\author{Vandiver Chaplin}
\affiliation{University of Alabama in Huntsville, Huntsville, AL 35899, USA}
\author{Elif Beklen}
\affiliation{Physics Department, Suleyman Demirel University, 32260 Isparta, Turkey}
\affiliation{Max-Planck Institut fur Extraterrestische Physik, 85748, Garching, Germany}
\author{Mark Finger}
\affiliation{Universities Space Research Association, Huntsville, AL 35805, USA}
\author{Narayana
Bhat}
\affiliation{University of Alabama in Huntsville, Huntsville, AL 35899, USA}
\author{Michael S. Briggs}
\affiliation{University of Alabama in Huntsville, Huntsville, AL 35899, USA}
\author{Valerie Connaughtont}
\affiliation{University of Alabama in Huntsville, Huntsville, AL 35899, USA}
\author{Jochen Greiner}
\affiliation{Max-Planck Institut f¬ur Extraterrestische Physik, 85748, Garching, Germany}
\author{R. Marc Kippen}
\affiliation{Los Alamos National Laboratory, Los Alamos, NM 87545}
\author{Charles A. Meegant}
\affiliation{Universities Space Research Association, Huntsville, AL 35805, USA}
\author{William S. Paciesas}
\affiliation{Universities Space Research Association, Huntsville, AL 35805, USA}
\author{Robert Preece}
\affiliation{University of Alabama in Huntsville, Huntsville, AL 35899, USA}
\author{Andreas von Kienlin}
\affiliation{Max-Planck Institut f¬ur Extraterrestische Physik, 85748, Garching, Germany}

\begin{abstract}
The Gamma ray Burst Monitor (GBM) on board Fermi Gamma-ray Space Telescope has been providing continuous data to the astronomical community since 2008 August 12.  We will present the results of the analysis of the first three years of these continuous data using the Earth occultation technique to monitor a catalog of 209 sources. Although the occultation technique is in principle quite simple, in practice there are many complications including the dynamic instrument response, source confusion, and scattering in the Earth's atmosphere, which will be described.  We detect 99 sources, including 40 low-mass X-ray binary/neutron star systems, 31 high-mass X-ray binary/neutron star systems, 12 black hole binaries, 12 active galaxies, 2 other sources, plus the Crab Nebula and the Sun. Nine of these sources are detected in the 100-300 keV band, including seven black-hole binaries, the active galaxy Cen A, and the Crab. The Crab and Cyg X-1 are also detected in the 300-500 keV band. GBM provides complementary data to other sky monitors below 100 keV and is the only all-sky monitor above 100 keV.  In our fourth year of monitoring, we have already increased the number of transient sources detected and expect several of the weaker persistent sources to cross the detection threshold.  I will briefly discuss these new sources and what to expect from our five year occultation catalog.
\end{abstract}

\maketitle

\thispagestyle{fancy}

\section{Introduction}
The low energy gamma-ray/hard X-ray sky is populated largely by active X-ray
binaries, active galactic nuclei, soft gamma ray repeaters, the Crab, and the Sun.  The Earth occultation technique (EOT) produces flux measurements of individual source with non-imaging detectors  by detecting the step-like feature
in the counting rate as the source passes into or moves out of eclipse by the Earth. Here
we present a results of the analysis of sources over the energy range 8 - 1000 keV utilizing EOT with the set of sodium iodide detectors aboard the Gamma-ray Burst Monitor
(GBM) instrument \cite{Meegan_2009} on \emph{Fermi}.
\subsection{GBM}
	The Fermi Gamma-ray Burst Monitor (GBM) consists of 14 detectors: 12 NaI
detectors, each 12.7 cm in diameter and 1.27 cm thick (each with effective area $\sim123$ cm$^{2}$
at 100 keV); and two BGO detectors, 12.7 cm in diameter and 12.7 cm thick (each with
effective area $\sim 120$ cm$^{2}$ in the 0.15Ð2 MeV range). The NaI detectors are located on the
corners of the spacecraft, with different orientations in order to  provide nearly uniform coverage of the unocculted sky in the
energy range from 8 keV to 1 MeV. Typically 3-4 NaI detectors view an Earth occultation
within 60 degrees of the detector normal vector. The two BGO detectors are located on
opposite sides of the spacecraft and view a large part of the sky in the energy range $\sim150$
keV to $\sim40$ MeV.
	
	GBM has two continuous data types: CTIME data with nominal 0.256-second time
resolution and 8-channel spectral resolution and CSPEC data with nominal 4.096-second
time resolution and 128-channel spectral resolution. The results presented here, with the exception of spectral analysis, use the lower-spectral resolution CTIME data for the NaI detectors. 
%


\subsection{Step Fitting Techniques}
\begin{figure}[t]
\centering
\includegraphics[width=70mm]{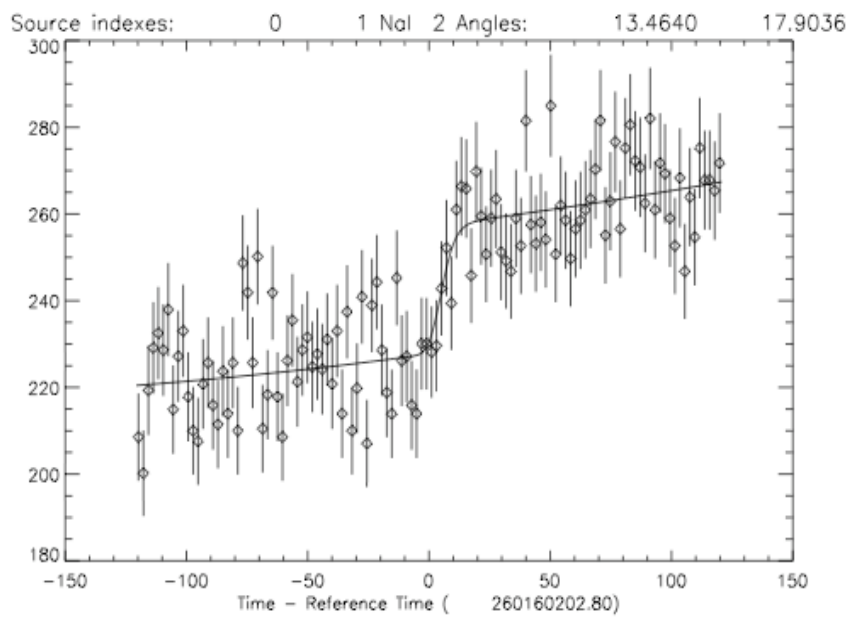}
\caption{\footnotesize{Example of an occultation fit window.  The occultation step for the source is clearly seen at t=0.  The best model for the fit window is shown by the solid curve.}} \label{step}
\end{figure}

We have adapted the technique of Harmon et al. (2002) for GBM. This technique
involves fitting a model consisting of a quadratic background plus source terms to a short
($\sim4$ min) window of data centered on the occultation time of the source of interest (See Figure \ref{step}) . For
GBM we have incorporated the changing detector response across the fit window into our
source terms. 
Before any sources are fitted, good time intervals (GTIs) of GBM data are defined.
The GTI intervals from the CTIME data files are shortened by 10 s to remove transient
events due to GBM high voltage turn-on and turn-off. A spline model is fitted to the 12-25
keV CTIME data to eliminate large background deviations, typically on $\sim100$s time scales,
due to South Atlantic Anomaly entries and exits, bright solar flares, gamma-ray bursts, and
other brief bright events from the GTIs.

For each day, the occultation times for each source in the catalog are calculated using
the known spacecraft positions. The time of each occultation step is taken to be the time
for which the transmission of a 100 keV gamma ray through the atmospheric column is
50\%. The time at which the atmospheric transmission reaches 50\% is energy dependent,
with lower energies absorbed at lower atmospheric densities so that a setting step will occur
earlier than at higher energies.

Measuring the flux from the source of interest requires fitting a model of all sources within the fit window plus a quadratic background function to the count
rate data for each detector and each energy channel.  The model includes the detector responses, assumed energy spectrum and
atmospheric transmission.  The fit results in a weighted scale factor for all detectors that is used to estimate fluxes;
 \begin{equation}\label{eq:units}
F(E_{ch}) = \overline{a}(E_{ch})*\int_{E_{ph}} f(E_{ph})dE_{ph}
\end{equation}
where $\overline{a}(E_{ch})$ is the wighted mean of scale factors for each detector and $f(E_{ph})$ is the flux model for each source in the fit window.
\subsection{Systematic Effects}
From our experience using the Earth occultation technique with BATSE and with
GBM, we have identified the following systematic effects that affect Earth occultation flux
measurements: (1) accuracy of the assumed source spectra, (2) large variations in the
background, (3) duration of the occultation transition, (4) inaccuracies in the detector
response matrices, (5) occultation limb geometry, and (6) nearby sources. In this section we
describe our mitigation strategies and our efforts to reduce, account for, or quantify these
effects.

By significantly deviating from the canonical spectrum for well known sources, we explored the systematic effects due to the accuracy of the source model.  The detection significance was unaffected by the assumed spectra; however, the flux values were somewhat dependent on source model especially if a hard source model was used for an especially soft source or vice versa.  Other sources in the fit window were not affected.

Because the background and sources are fit simultaneously, variations in the background
that cannot be fitted with a quadratic function will be absorbed into the source terms
and can potentially affect source measurements.  Occultation steps when the spacecraft spin rate exceeds $4\times10^{-3}$rad s$^{-1}$ are flagged and excluded from analysis.  In addition, GTIs exclude large variations in the background.

If a source is at a high latitude, the occultation will happen very slowly with respect to the window duration and step fitting becomes difficult.  Steps in which the occultation exceeds 20 seconds are flagged and excluded from analysis.

 The most noticeable systematic effect arises from the detector response matrices
when solar panel occultations of sources. In the mass model for Fermi used to derive the
responses, the solar panels are included at a fixed orientation.  The true orientation of the solar panels is not fixed and can cause an unmodeled blockage of a detector.  Detectors, in which any orientation of the solar panels may block the source, are excluded from the step fitting.

Every $\sim52$ days, the Fermi orbit precesses so that the Earth occultation limb
geometry (i.e. the projection of the EarthÕs limb on the sky) repeats at this period. For a particular geometry, Earth occultation flux measurements may be systematically low or
high due to unmodelled sources or non-point source backgrounds such as galactic ridge
emission. These systematic effects can be very difficult to identify. To mitigate these
effects, we use \emph{Swift}/Burst Alert Telescope (\emph{Swift}/BAT) data to populate a flare database listing times and brightness
levels for potentially interfering sources.  The flare database along with sources identified using Earth occultation imaging \cite{Rodi_2013} are used to identify the sources used in the step fitting. 

Nearby sources, especially bright or highly variable sources, can make occultation
measurements very difficult. If two occultation steps occur within 8 seconds of each other,
the step fitting breaks down, so we automatically flag these steps and exclude them from
further analysis.  Extraordinary event such as the bright outburst of A0525+26 in December 2009 will affect nearby sources such as the Crab.  These effects are easily identified and we manually flag the data and exclude them from further analysis.
\subsection{``Ghost'' source analysis}
To examine the remaining systematic effects, 3-year light curves were selected from
512 ``ghost'' sources run through the occultation software in conjunction with GBM
imaging analyses (Rodi et al. 2011). The initial list was reduced by excluding any ``ghost''
sources within $\pm10$ degrees in longitude and latitude of the galactic center. Any ``ghost'' sources
within 2 degrees of a source in the GBM catalog were also removed, resulting in a sample of
about 200 ``ghost'' sources distributed over the whole sky. We investigated scatter plots of
various combinations of parameters including location on the sky, flux, flux significance,
flux statistical error, and flux standard deviation in all eight energy bands.


These distributions were used to estimate overall systematic errors on the flux
measurements. Since no source was expected at these locations, the distribution of the flux
significance in each channel is expected to be centered on zero with a standard distribution
of 1.0 (See \cite{Wilson-Hodge_2012} for details).  Estimates of the systematic errors derived from this analysis are shown
in Table \ref{table}.

\begin{table}[t]
\begin{center}
\caption{Systematic error estimates for GBM Earth occultation analysis }
\begin{tabular}{lc}
\hline
\hline
\label{table}
Energy band \ \ \ \ \ \ \ \ & Systematic error \\
 (keV) & (mCrab)\\
\hline
8-12 & 3.4 \\
12-25 & 2.8 \\
25-50 & 2.2  \\
50-100 & 1.5 \\
100-300 & 3.1 \\
300-500 & 3.4 \\
\hline
\end{tabular}
\end{center}
\end{table}
\subsection{Sensitivity}
\begin{figure}[t]
\centering
\includegraphics[width=70mm]{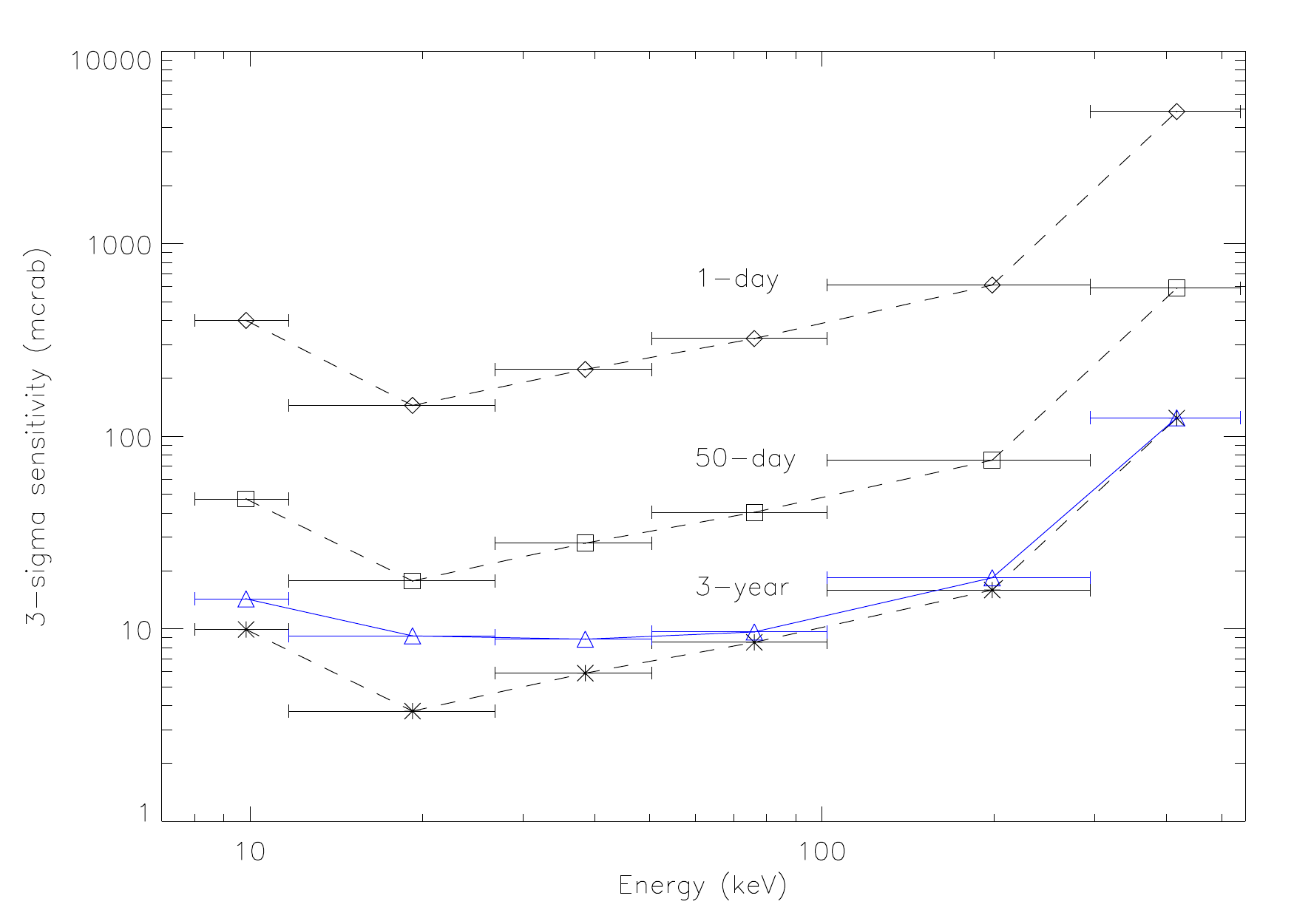}
\caption{\footnotesize{Estimated 3-$\sigma$ sensitivities for the GBM Earth occultation technique from 8-500
keV. Statistical estimates are shown for 1-day (diamonds), 50-day (squares), and 3-year
(asterisks) intervals. Systematic errors from Table 1 have been added in quadrature to the
3-year statistical errors and are plotted as triangles connected with solid lines.}} \label{sensitivity}
\end{figure}

Calculating the sensitivity for the Earth occultation technique is made challenging by
the constantly changing spacecraft source geometry, detector response, and the constantly
changing hard X-ray sky. To estimate the statistical sensitivity of the technique, we
used four detected sources, the Crab (a galactic source in a relatively uncrowded region),
Centaurus A (an active galaxy at high galactic latitude and high declination to account for
steps lost when the source was at high beta angles), NGC 4151 (another active galaxy at
high galactic latitude and moderate declination), and GRS 1915+105 (a persistent black
hole system near but not at the galactic center). For each of these sources, 1-day, 50-day,
and 3-year average measured fluxes and statistical errors were computed. The average error
in each energy channel for each source was then computed. The approximate statistical
sensitivity for each source was computed as three times the average error in mCrab
units. For the three year averages only the systematic errors from Table \ref{table} were added in
quadrature. Figure \ref{sensitivity} shows the estimated sensitivities, averaged for the four sources.

\subsection{Comparisons}
As a final check we compared our occultation results in the energy range from 12-50 keV to \emph{Swift}/BAT 15-50 keV survey results for eight sources using 2-4 day averages. 
  We obtained very close agreement with the exception of Sco X-1.  The excess in Sco X-1 was expected due to the particularly soft spectrum and the lower energy band used for GBM occultation analysis.  
\section{Results}
\begin{figure}[t]
\centering
\includegraphics[width=80mm]{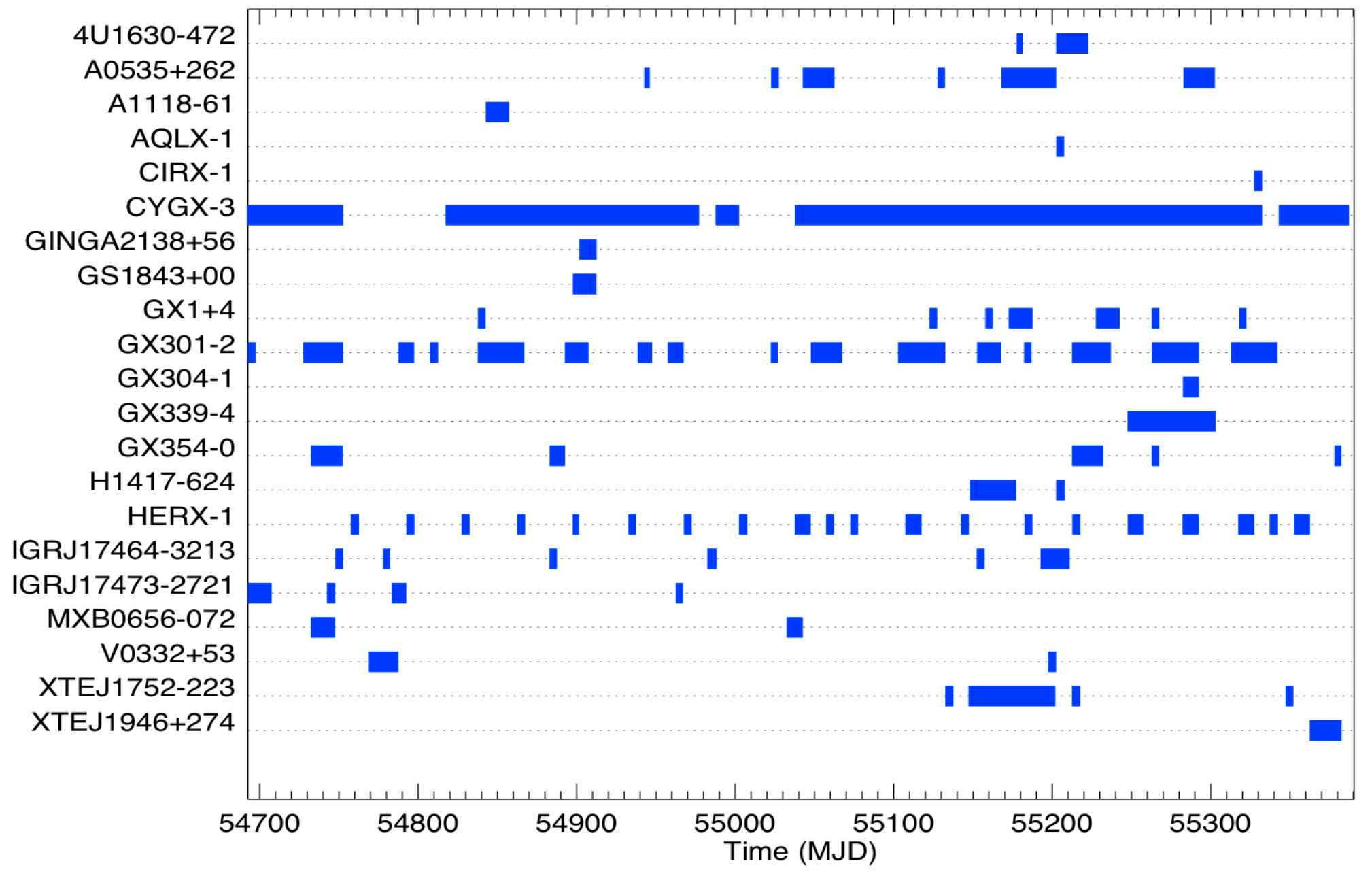}
\caption{\footnotesize{A sample of transients detected by the Earth occultation method.  The solid bars indicate the times in which the transient was detected at the $5\sigma$  level.}}\label{transients}
\end{figure}

In August 2012 we published our 3 year catalog \cite{Wilson-Hodge_2012} that included 209 sources.  99 sources were detected at $5\sigma$ (Category A) including 40 LMXB/NS, 31 HMXB/NS, 12 BHC, 12 AGN, and the Crab Nebula.  We also had several sources detected between 3.5 and $5\sigma$ (Category B).  We continue to add sources to our catalog, currently 
219 sources, and additional transients are being detected regularly (See Figure \ref{transients}).

\subsection{Sources above 100 keV}
One of EOT's special abilities is to monitor sources above 100 keV.  We regularly detect 4 black hole candidates (BHC); Cygnus X-1, 1E 1740-29, GRS 1915+105, and Swift J1753.5-0127, the Seyfert 2 galaxy Cen A  and the Crab Nebula above 100 keV.  We have also detected flares in the BHCs XTE J1752-223 and GX 339-4 above 100 keV.  
\subsubsection{Crab Nebula}
\begin{figure}[t]
\centering
\includegraphics[width=80mm]{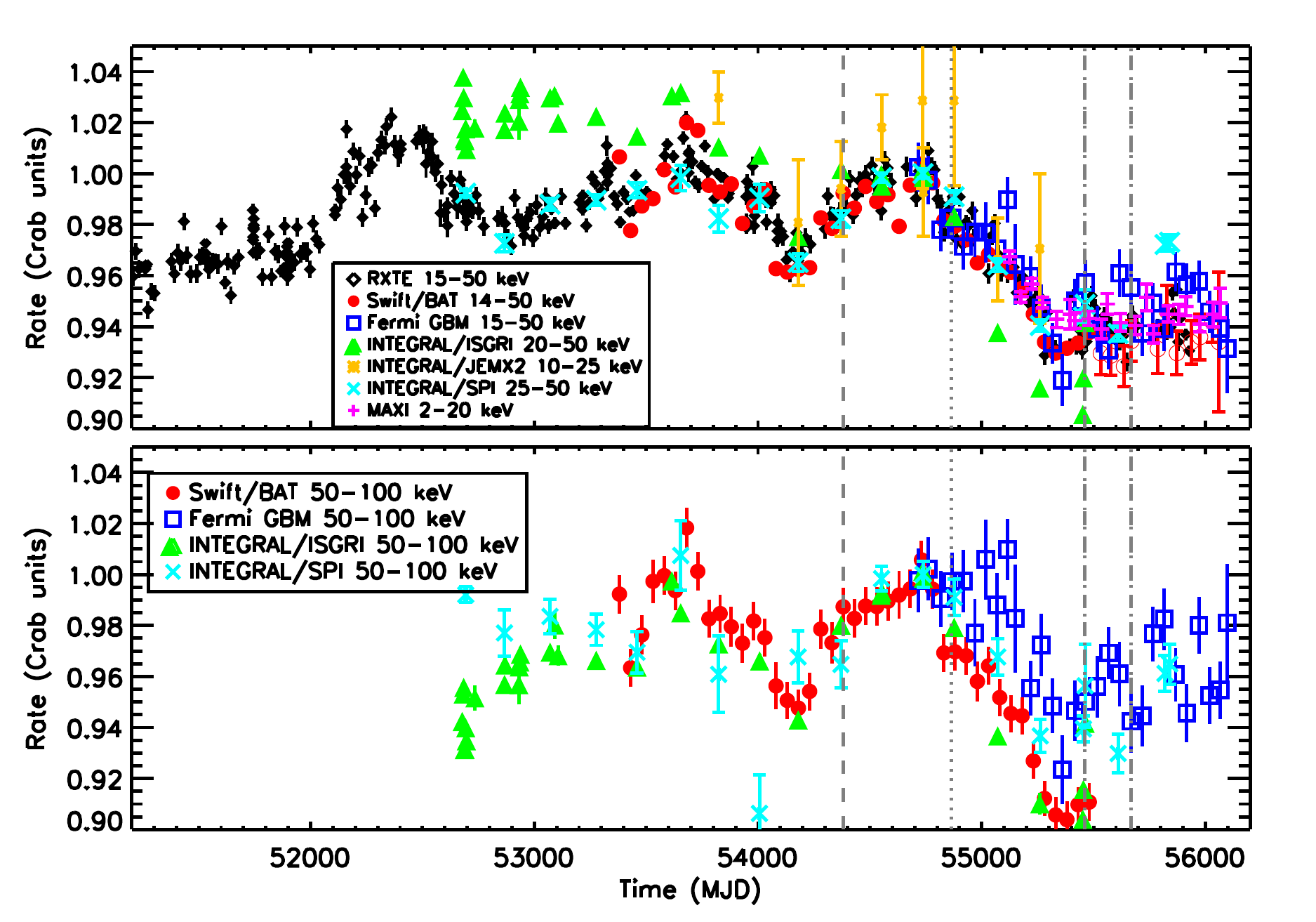}
\caption{\footnotesize{Top - A light curve of the Crab Nebula from 15-50 keV from \emph{Fermi}/GBM (blue squares), RXTE (black diamonds), \emph{Swift}/BAT (red filled circles), INTEGRAL/ISGRI, INTEGRAL/JEMX2, INTEGRAL/SPI, (green triangles, yellow squares, blue X's respectively) and MAXI 2-20 keV (purple pluses).  Rates are normalized to \emph{Fermi}/GBM measurements in 2008 August.   Bottom - A similar light curve of the Crab Nebula from 50-100 keV from \emph{Fermi}/GBM (blue squares), RXTE (black diamonds), \emph{Swift}/BAT (red filled circles), INTEGRAL/ISGRI (green triangles), and INTEGRAL/SPI (blue X's). }}\label{crab}
\end{figure}

The Crab Nebula is detected regularly up to 300 keV in our monitoring program.  Since 2008 August, a $\sim7.0\%$ (70 mCrab) decline has been observed in the overall Crab Nebula flux in the 15-50 keV band \cite{Wilson-Hodge_2011}.  This decline was independently confirmed in the 15-50 keV band with \emph{Swift}/BAT, the \emph{Rossi X-ray Timing Explorer} Proportional Counter Array (\emph{RXTE}/PCA), and the Imager on-Board the \emph{INTEGRAL} Satellite (IBIS) (See Figure \ref{crab}).  This was the first conformation of the X-ray variability of the Crab Nebula.

\begin{figure}[t]
 \centering
\includegraphics[width=80mm]{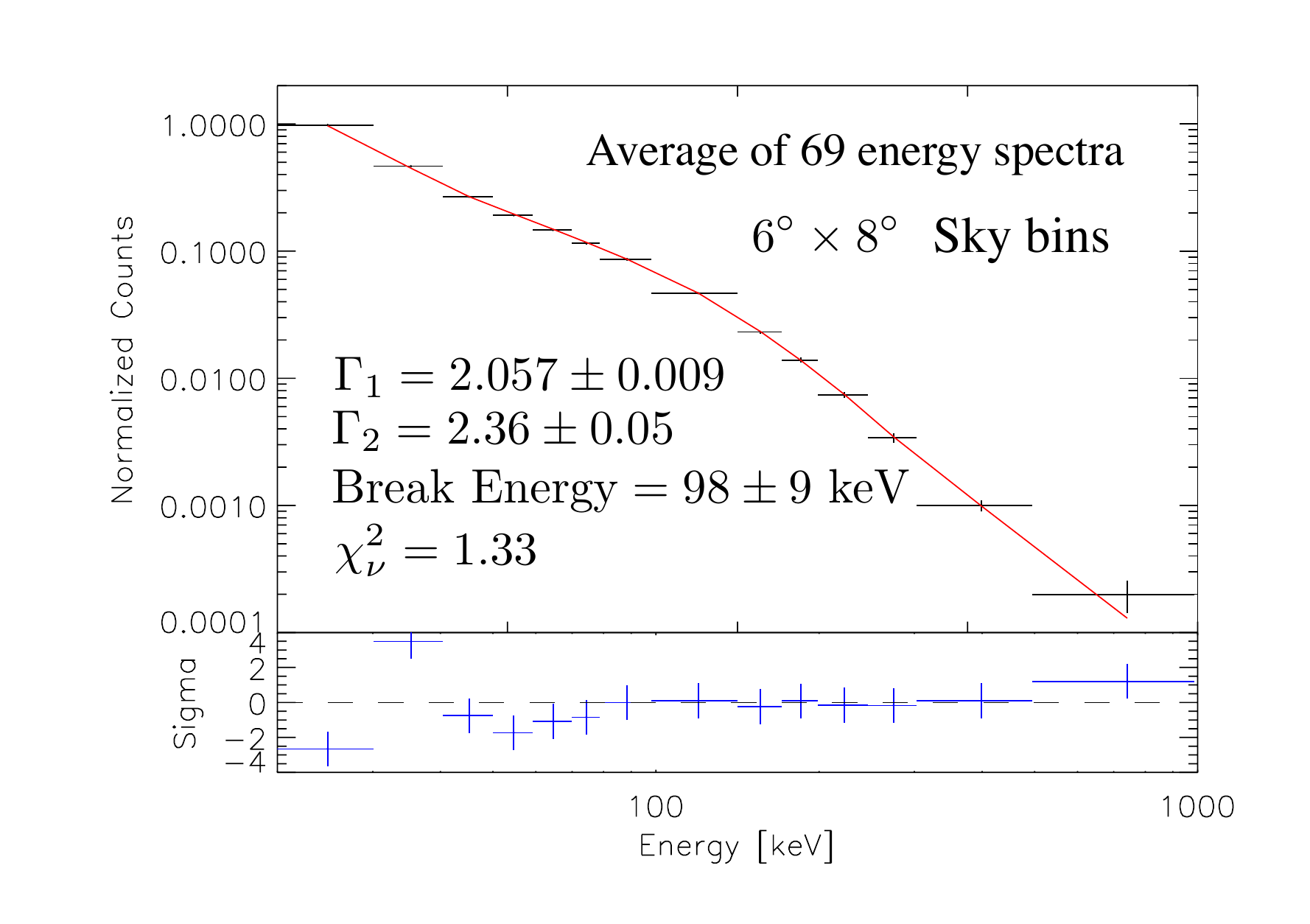}
\caption{\footnotesize{Points are the average of 69 spectra for the Crab while the red curve is the best fit model.}}
\label{spectrum}
\end{figure}

	The EOT is also capable of spectral analysis by binning the sky into instrument response bins in spacecraft coordinates.  By using CSPEC data we customize our energy channels to best match the source of interest.  Figure \ref{spectrum} shows an average 69 energy spectra from the Crab utilizing 14,629 occultation steps for an exposure time of approximately 1.7Ms (assuming 120s of exposure per step).  The spectrum was fit with a broken power-law model and the best fit parameters were $\Gamma_{1} = 2.057\pm0.009, \Gamma_{2} = 2.36 \pm0.05$ with the break at 98$\pm$9 keV and a $\chi^{2} = 1280.44$ with 963 degrees of freedom ($\chi^{2}_{\nu} = 1.33$).  This result is consistent with \cite{Jourdain_2009} who fit \emph{INTEGRAL}/SPI Crab data. 

\begin{figure}[t]
 \centering
\includegraphics[width=50mm]{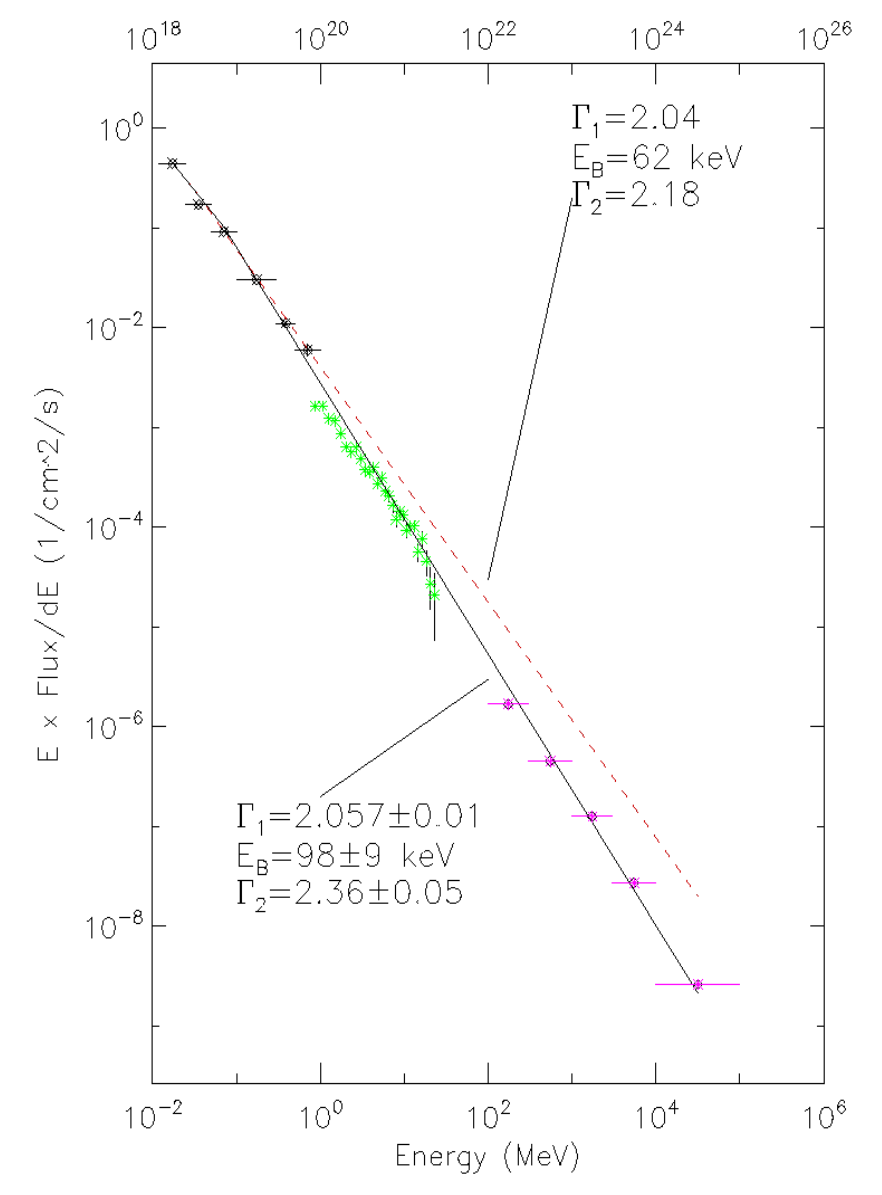}
\caption{\footnotesize{The black points are the average of 69 spectra obtained by the EOT.  The green points are data obtained from COMPTEL, while the pink points are the data from the \emph{Fermi}/LAT 2FGL catalog.  The solid curve is the extrapolation of the best fit spectrum from the EOT while the dashed line is the extrapolation of the spectrum obtained by \cite{Jourdain_2009}. }}\label{spectrum2}
\end{figure}

Extrapolating the Crab spectrum obtained from GBM EOT to GeV energies obtained from the \emph{Fermi}/LAT 2FGL catalog suggests that the power-law is an adequate description of the data  through GeV energies (See Figure \ref{spectrum2}).

\subsubsection{Cygnus X-1}
\begin{figure}[t]
 \centering
\includegraphics[width=70mm]{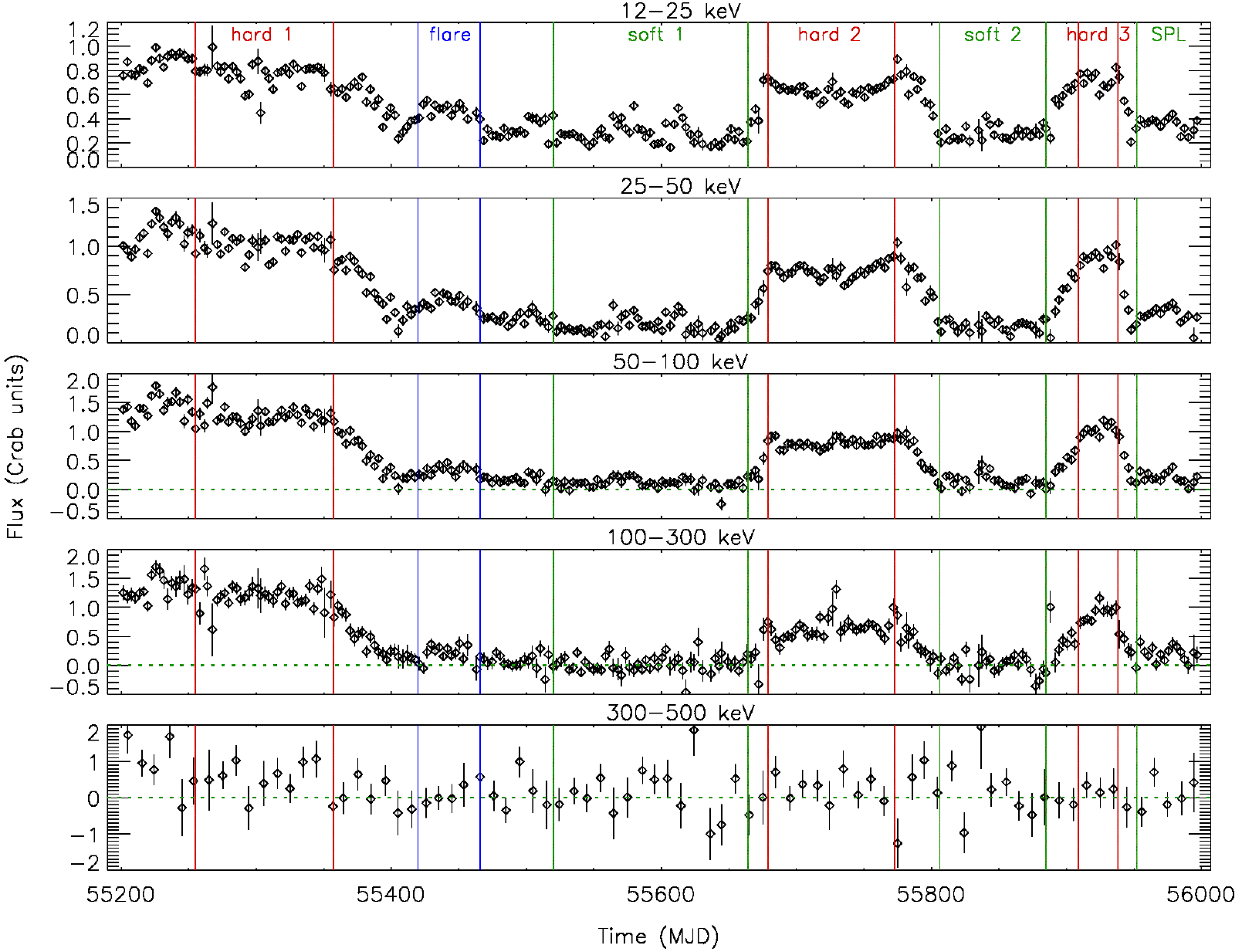}
\caption{\footnotesize{The EOT measured fluxes for Cygnus X-1 shown for 12-25, 25-50, 50-100, 100-300, 300-500 keV energy bands.  The vertical colored lines indicate times of state changes. } }\label{cygx-1}
\end{figure}

The BHC Cygnus X-1 is regularly observed with EOT to 300 keV especially during it's hard/low state.  GBM's EOT has routinely identified state changes in this black hole candidate \cite{Case_2010}.  Figure \ref{cygx-1}  shows light curves for Cygnus X-1 from 12 - 500 keV where the state changes and flares are readily seen seen up to 300 keV.

\subsubsection{XTE J1752-223}
\begin{figure}[t]
   \centering
   \includegraphics[width=60mm]{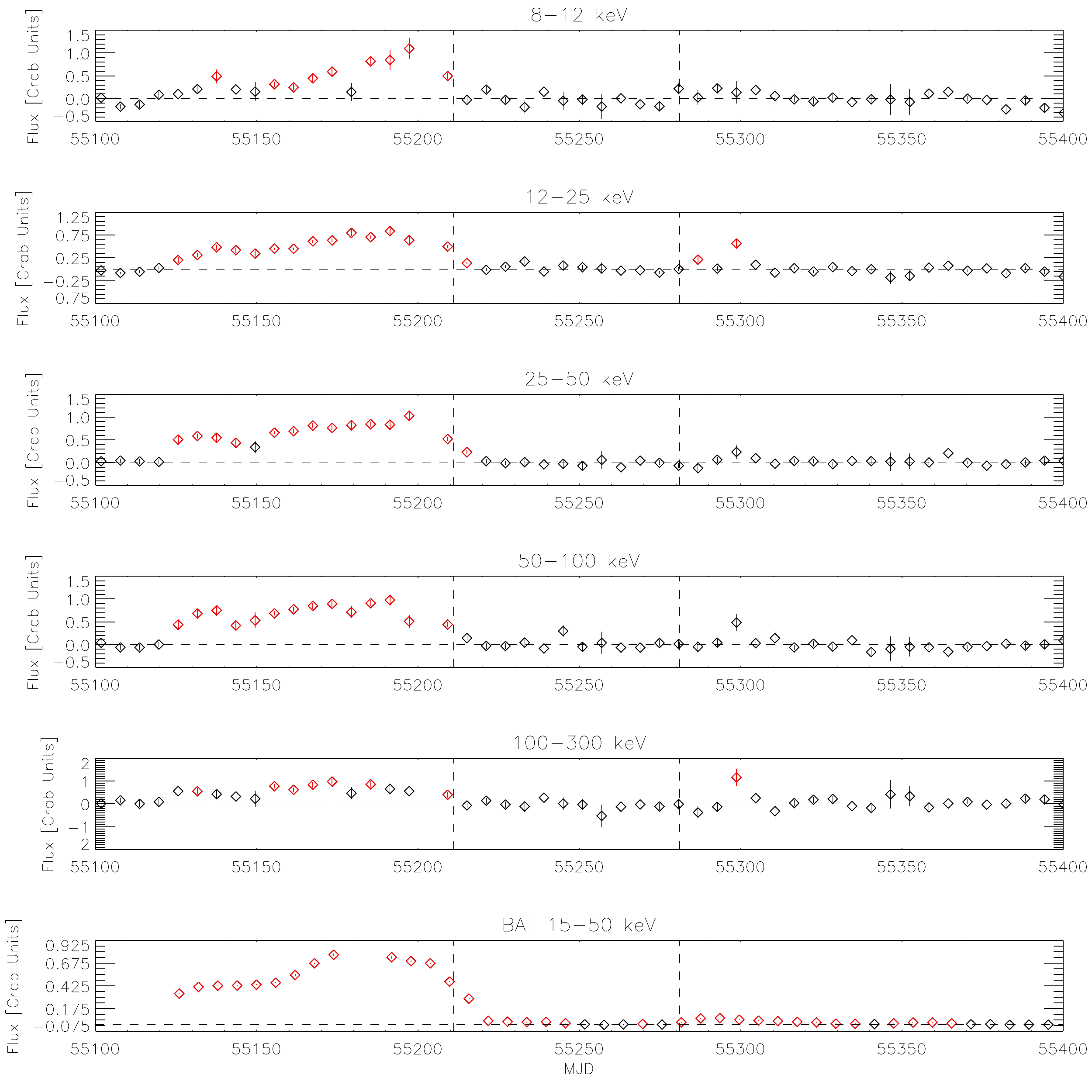}
   \caption{\footnotesize{Light curves of XTE J1752-223, from the GBM occultation project, in five energy bands (8-12 keV, 12-25 keV, 25-50 keV, 50-100 keV, and 100-300 keV) integrated over 6 days.  The bottom panel shows the Swift/BAT 15-50 keV survey data for the same period.  Red points denote 3$\sigma$ detections or better.  The vertical dashed lines indicate state changes from hard to soft and again from soft to hard.}}
   \label{fig:lc}
\end{figure}

The transient BHC XTE J1752-223 was discovered by RXTE \cite{Markwardt_2009} on 2009 October 23.
Fermi/GBM, using the EOT, observed XTE J1752-223 from October 23, 2009 (MJD 55127) to January 20, 2010 (MJD 55216) \cite{Wilson-Hodge_2009} while the source was in a spectrally hard state.  The EOT stopped detecting the source when the source shifted from a spectrally hard to soft state \cite{Curran_2011}.  Fermi/GBM again detected the source between April 1, 2010 (MJD 55287) to April 13, 2010 (MJD 55299) when the source again transitioned to a hard state before its emission fell back to quiescence.  Figure \ref{fig:lc} shows the light curves, from the GBM occultation project, in five energy bands (8-12 keV, 12-25 keV, 25-50 keV, 50-100 keV, and 100-300 keV).  The figure also shows the Swift/BAT 15-50 keV survey data for the same period.

\subsubsection{GX 339-4}
\begin{figure}[t]
   \centering
   \includegraphics[width=60mm]{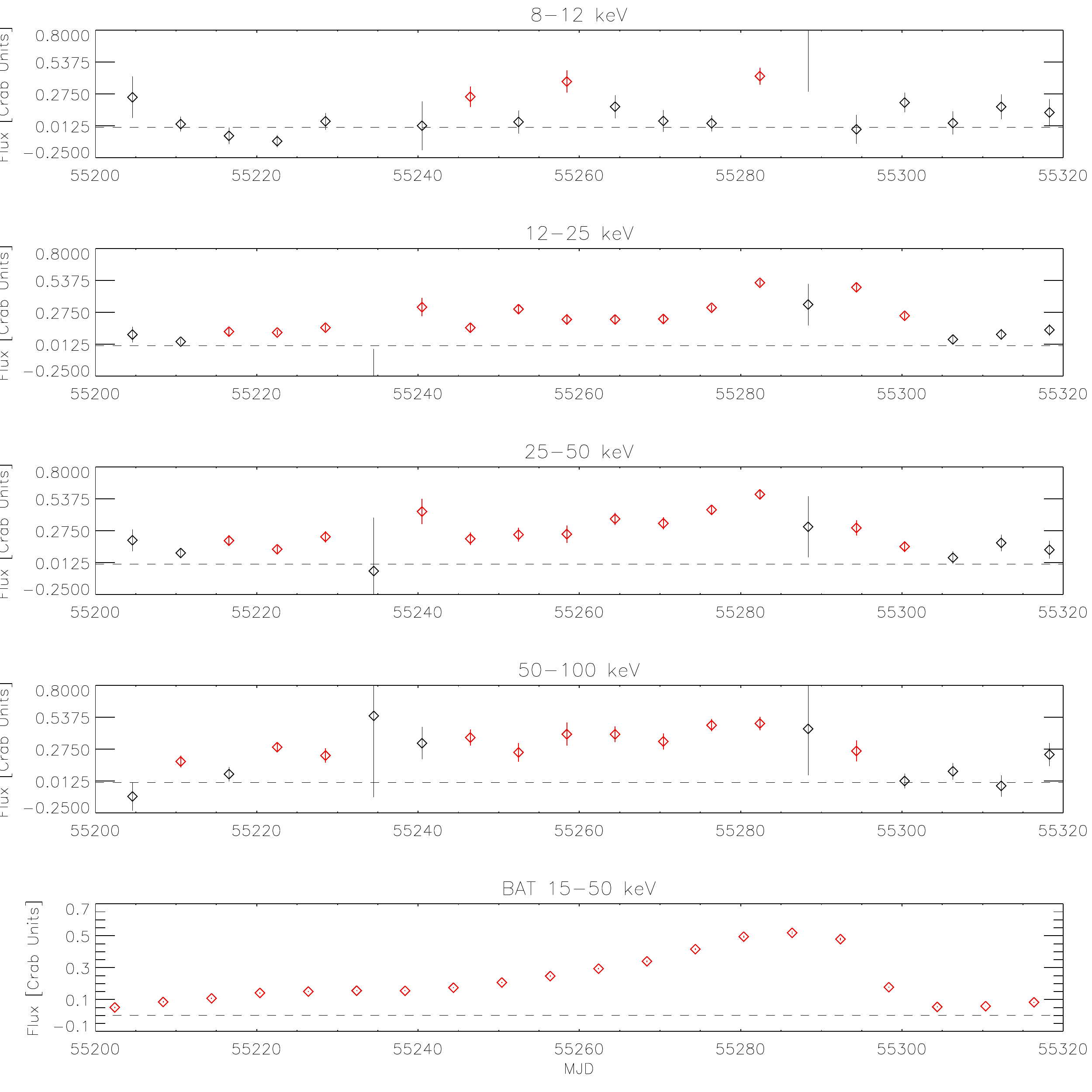}
   \caption{\footnotesize{Light curves of GX 339-4 from the GBM occultation project, in four energy bands (8-12 keV, 12-25 keV, 25-50 keV and 50-100 keV) integrated over 6 days.  The bottom panel shows the Swift/BAT 15-50 keV survey data for the same period.  Red points denote 3$\sigma$ detections or better.  The vertical dashed lines indicate state changes from hard to soft and again from soft to hard.}}
   \label{gx339}
\end{figure}

The BHC GX 339-4 is characterized by rapid time variability and low/hard X-ray states similar to Cygnus X-1.
The flux observed by GBM began to increase starting in early 2010 January \cite{Case_2011} and continued to increase up to a level of $\sim400$ mCrab (12Ð25 keV), $\sim650$ mCrab (25Ð50 keV), $\sim800$ mCrab (50Ð100 keV), and $\sim550$ mCrab (100Ð300 keV) by early 2010 April, after which it began to rapidly decrease. It returned to quiescence in the higher energy bands by mid-April and in the 12Ð50 keV band by the end of April.  Figure \ref{gx339} shows the light curves, from the GBM occultation project, in four energy bands (8-12 keV, 12-25 keV, 25-50 keV and 50-100 keV).  The figure also shows the Swift/BAT 15-50 keV survey data for the same period.

\subsection{Additional Results - A 0535+26}
GBM EOT continues to provide valuable information on BHCs and on other sources such as Be X-ray binaries as well as active galaxies in the energies between 8-1000 keV.  Monitoring sources for outbursts and state changes provides valuable information to the X-ray and gamma-ray community.  	

\begin{figure}[t]
   \centering
   \includegraphics[width=60mm]{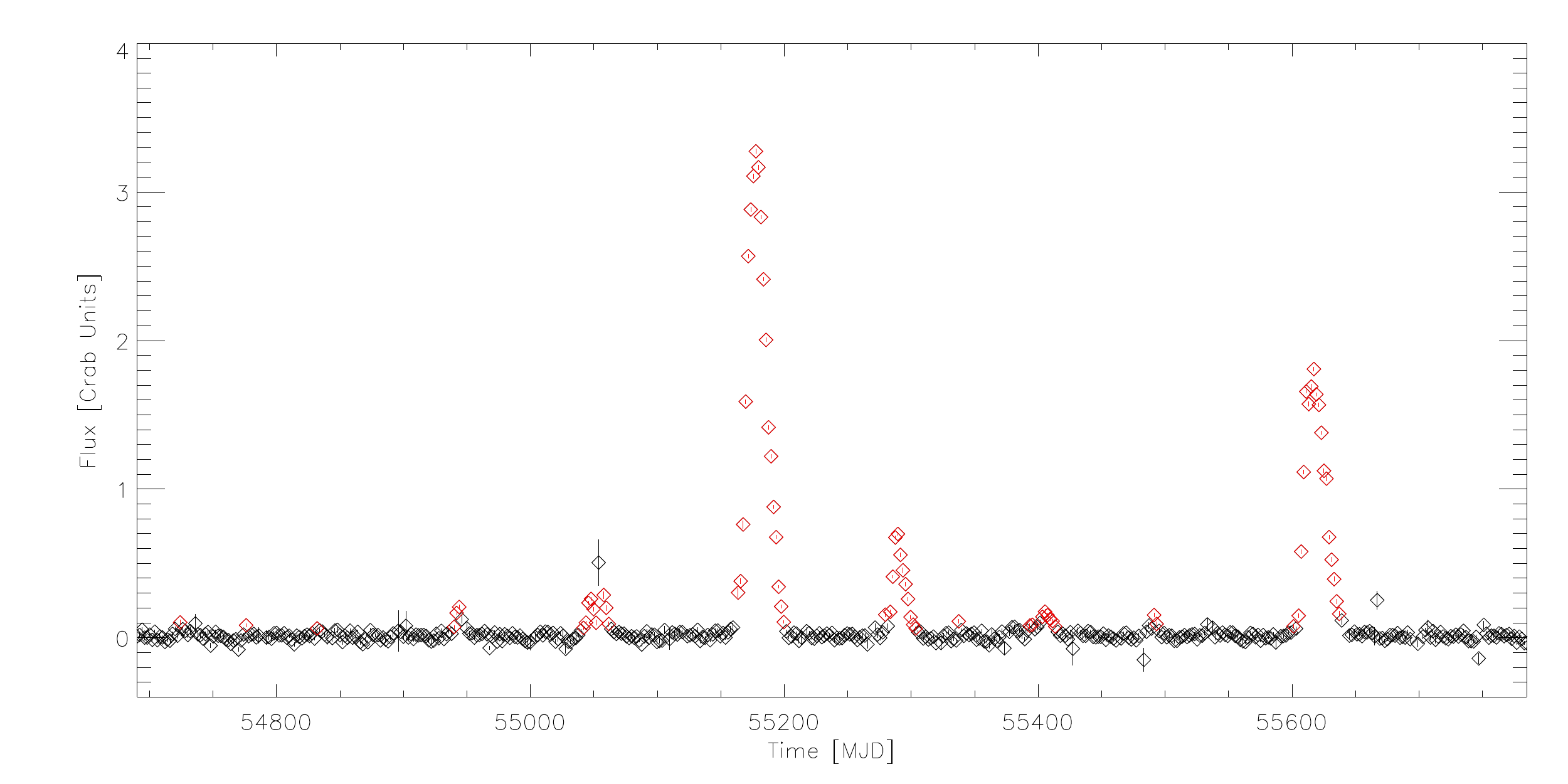}
   \caption{\footnotesize{Two day average fluxes for A0535+26, from the GBM occultation project in the energy from 12-100 keV.  The four normal outbursts prior to the giant outburst in 2009 are detected as well as the following three normal outburst as well as the second giant outburst in 2012.  The red points are $5\sigma$ detections or better}}
   \label{A0535}
\end{figure}

A 0535+26 is a Be X-ray binary that exhibits transient outbursts occasionally associated with periastron passage.  Generally two types of outbursts are observed, smaller ones that peak at a few hundred mCrabs and giant outbursts that may peak at several Crab.  As we learn more about these outbursts, it becomes more difficult to separate the two classifications \cite{Choni_2012}.  In 2010 April an intermediate outburst $\sim1$ Crab was observed (See Figure \ref{A0535}).
\section{Conclusion}
GBM EOT currently provides online light curves and data for 218 sources.  We are currently automating our spectral capabilities and will soon have spectra available on our webpage\footnote{http://heastro.phys.lsu.edu/gbm/}.  New sources are constantly being added to our monitoring program and we expect to significantly increase the number of sources detected in our next catalog.  In addition, we expect to role out new data products that promise to improve the versatility of the analysis currently available.  In sum, the EOT project is a dynamic project that keeps improving and expanding to the needs of the scientific community. 
\vspace{-0.76cm}
\vspace{-0.15cm}
\bibliography{Fermi2012_Pete_short}

\end{document}